# Modulation of ionic current rectification in short bipolar nanopores

Hongwen Zhang,[1,2] Long Ma,[1] Chao Zhang,[3] and Yinghua Qiu[1,2,4*]

1. Shenzhen Research Institute of Shandong University, Shenzhen, 518000, China

2. Key Laboratory of High Efficiency and Clean Mechanical Manufacture of Ministry of Education, National Demonstration Center for Experimental Mechanical Engineering Education, School of Mechanical Engineering, Shandong University, Jinan, 250061, China

3. School of Mechanical and Electronic Engineering, Shandong Jianzhu University, Jinan, 250101, China

4. Suzhou Research Institute of Shandong University, Suzhou, 215123, China

*Corresponding author: yinghua.qiu@sdu.edu.cn

## Abstract

Bipolar nanopores, with asymmetric charge distributions, can induce significant ionic current rectification (ICR) at ultra-short lengths, finding potential applications in nanofluidic devices, energy conversion, and other related fields. Here, with simulations, we investigated the characteristics of ion transport and modulation of ICR inside bipolar nanopores. With bipolar nanopores of half-positive and half-negative surfaces, the most significant ICR phenomenon appears at various concentrations. In these cases, ICR ratios are independent of electrolyte types. In other cases where nanopores have oppositely charged surfaces in different lengths, ICR ratios are related to the mobility of anions and cations. The pore length and surface charge density can enhance ICR. As the pore length increases, ICR ratios first increase and then approach their saturation which is determined by the surface charge density. External surface charges of nanopores can promote the ICR phenomenon mainly due to the enhancement of ion enrichment inside nanopores by external surface conductance. The effective width of exterior charged surfaces under various conditions is also explored, which is inversely proportional to the pore length and salt concentration, and linearly related to the pore diameter, surface charge density, and applied voltage. Our results may provide guidance for the design of bipolar porous membranes.

**TOC**

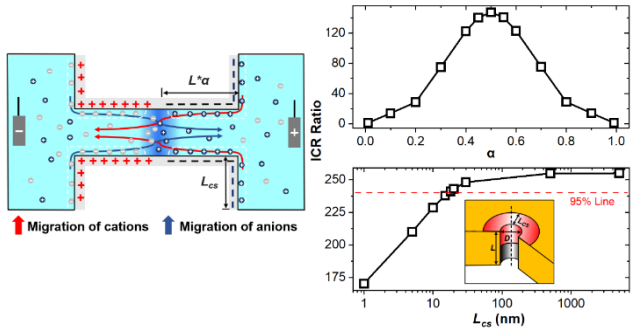

# Introduction

Nanopores provide a versatile platform for the investigation of ion and fluid transport under confined spaces,[1, 2] which has attracted tremendous research attention in various applications, such as nanofluidics,[3, 4] desalination,[5, 6] bio-sensing,[7] and energy conversion.[8, 9] In such highly confined spaces, the wall properties of nanopores exhibit significant modulation of the behavior of fluid and ion transport. For example, the hydrophobicity of pore walls can play a gating role in fluid transport through nanopores.[10, 11] Surface charges can attract counterions and repel co-ions in the solution which enables the ionic selectivity of nanopores.[12, 13] The induced electric double layers near charged pore walls can facilitate the migration and diffusion of counterions through nanopores.[14-17]

In aqueous solutions, cylindrical nanopores exhibit linear current-voltage curves. Uniform surface charges on pore walls can promote the transport of counterions due to the surface conductance in EDLs.[14, 16] In cylindrical nanopores with asymmetrical surface charge distributions including unipolar and bipolar distributions,[18, 19] because of the modulation of surface charges, rectified ionic currents can be obtained with different current values appearing at equal voltages but different polarities.[13, 20, 21] This current characteristic similar to that in diodes is called ionic current rectification (ICR). ICR has been widely studied in conical nanopores,[21, 22] whose appearance is mainly due to the ion selectivity of the nanopores at the pore tip.[23] The induced ion enrichment and depletion inside nanopores under forward and backward electric fields correspond to the "on" and "off" states with higher and lower currents, respectively.[13, 24, 25] From the literature, significant ICR usually appears in long nanopores.[13] This is mainly attributed to the enhanced ion selectivity by large charged pore surfaces[2, 12] and the increased space inside the pore for ionic enrichment and depletion under opposite biases.[23, 24] For short nanopores, exterior surface charges can promote the ICR phenomenon due to the enhanced ion enrichment and depletion caused by the ion transport along exterior charged surfaces.[13, 16] Unipolar cylindrical nanopores with only a portion of charged inner-pore walls act as ionic diodes similar to charged conical nanopores, due to the ion selectivity from the charged parts.[26]

Bipolar nanopores carry both positive and negative surfaces. Different from charged conical nanopores and unipolar nanopores, the appearance of ICR in bipolar nanopores originates from the ionic selectivity from both charged pore ends.[18, 27] At the "on" state, cations and anions are selected respectively to the negatively and positively charged portion and ion enrichment forms inside nanopores. Under the opposite electric fields, the appeared ion depletion inside nanopores corresponds to the "off" state, due to the respective repulsion of cations and anions from positive and negative surface charges. Compared with conical nanopores and unipolar nanopores, bipolar nanopores can induce significant ICR with a small pore length.[27] With finite element simulations, Gracheva et al. investigated the ICR through a bipolar nanopore on a 24-nm-thick p-n semiconductor membrane.[28] Vlassiouk et al. compared ICR degrees in sub-20-nm-long bipolar nanopores with and without charged reservoirs.[27] In cases with charged reservoirs, obvious ICR appears in bipolar nanopores with even 2 nm in length. With molecular dynamic simulations, Luan et al.[29] built an ultra-thin diode in a bilayer h-BN membrane. Distinct ICR was observed via the modulation of ion transport from orifice charges. However, little systematic research on ICR in bipolar nanopores has been done in the literature under various situations, such as pore dimensions, surface charge properties, and electrolyte conditions. Also, from the previous work of our group and others,[13, 16, 27, 30-33] exterior surface charges may have a signification modulation of the ICR phenomenon through short bipolar nanopores. The effective charged area around the nanopore is still unknown, which serves as an important parameter in the design of porous membranes.[16, 17, 34]

Here, we systematically investigated the characteristics of ion transport through bipolar nanopores with COMSOL Multiphysics. With a series of simulation models, we explored the effect of the relative lengths of positive and negative surfaces on inner-pore walls on the ion transport characteristics. For bipolar nanopores with 100 nm in length, the most significant ICR phenomenon occurs in the case of half-positive and half-negative surfaces, where the ICR ratio is independent of the electrolyte type. For cases with different dimensions and surface charge properties of nanopores, as the pore length increases, the ICR ratio first increases and then reaches saturation, which value is proportional to the surface charge density. While the increase in the pore diameter reduces the ICR ratio. External surface charges significantly enhance the ICR ratio mainly due to the promotion of the ion enrichment inside nanopores. With the adjustment of the area of the charged exterior surface near the pore orifice, the effective width of the charged region is obtained, which is influenced by the pore parameters, such as the pore length, diameter, surface charge density, and length ratio of the negative surface, and simulation conditions, like the applied voltage, solution concentration, and electrolyte species. Due to the wide potential applications of bipolar nanopores in ionic current amplification,[18, 27] low-resistance conductance,[35] and osmotic energy conversion,[36-38] our results may be useful for the design and fabrication of bipolar nanoporous films.

## Simulation Methods

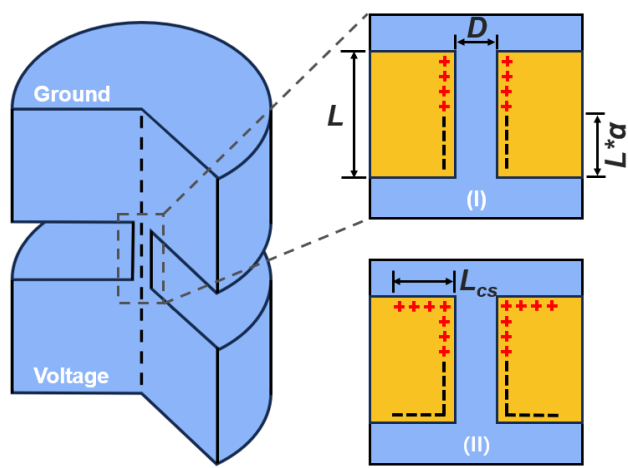

Figure 1 Simulation illustration of ion current through bipolar nanopores. The nanopore is located between two reservoirs. The length and diameter of the nanopore are marked as $L$ and $D$. Two simulation models are considered, i.e. the one with charged inner pore walls (model I), and the one with both charged inner and exterior pore walls (model II). The inner wall contains the distribution of bipolar charges, where α is the ratio of the negatively charged part to the total length (model I). When considering the effect of external surface charges on ion transport, the charged length of the external surface is set to $L_{cs}$ (model II).

Simulation models consisting of nanopores and two reservoirs were constructed using COMSOL Multiphysics to study the ion transport characteristics within the nanopores. As shown in Figure 1, two reservoirs with a radius and length of 5 μm were connected through the nanopore.[16, 17] In our simulations, the pore length ($L$) was changed from 5 to 1000 nm, and the pore diameter ($D$) was considered from 5 to 100 nm. The default pore length and diameter were 100 and 10 nm, respectively. Bipolar charge distributions were applied on inner-pore walls, with a portion carrying negative charges and the remaining portion carrying positive charges. The length ratio of the negatively charged part to the pore length was denoted as α, varying from 0 to 1 (Figure 1). In bipolar nanopores, the surface charge on both parts shares the same strength but different signs. The absolute value of the surface charge density changes from 0.02 to 0.08 $C/m^2$ with the default value of 0.08 $C/m^2$.[16, 39] Here, the selection of surface charge densities was based on different nanofluidic materials applied in experiments. For example, −0.005, −0.02, and −0.08 $C/m^2$ were reported for the surface charge density of SiN, $SiO_2$, and polyethylene terephthalate membranes, respectively.[39-43] In aqueous solutions, the surface charge density can be modulated by solution pH[44] and chemical modification.[45] In this work, the surface potential can be extracted from simulations and predicted with the Grahame equation.[44] Table S1 lists the values of surface potential obtained in the simulation cases. From our data, the surface potential from both methods share very close values.

Influences of charged exterior pore walls on ion transport in bipolar nanopores had also been considered, as shown by the simulation model II in Figure 1. The length of the charged exterior surface was denoted as $L_{cs}$, ranging from 0 to 5 µm.[16, 17] When $L_{cs}$ =0, the exterior surfaces were not charged, i.e. the simulation model I showed in Figure 1.

A series of simulations were conducted to investigate influences of surface charge distributions on nanopore walls and other parameters on the ion transport through nanopores, including different length ratios α, pore length and diameter, surface charge density, applied voltage, as well as salt type and concentration. NaCl, KCl, LiCl, and KF solutions were used to consider the influence of diffusion coefficients of cations and anions on the ion transport, of which KCl was the default electrolyte. The diffusion coefficients used for K, Na, Li, Cl, and F ions were $1.96×10^{-9}$, $1.33×10^{-9}$, $1.03×10^{-9}$, $2.03×10^{-9}$, and $1.47×10^{-9}$ m$^2$/s, respectively.[46] The salt concentration was changed from 10 to 500 mM with the default value of 100 mM. The applied voltage across the nanopore varied from −1 to 1 V, with the default value of 1 V. The system temperature remained constant at 298 K, and the dielectric constant of water was set to 80.

In our simulations, the Poisson-Nernst-Planck and Navier-Stokes equations (Eqs. 1-4) were coupled to simulate the detailed mass transport through nanopores, including the ion distributions at solid-liquid interfaces, as well as the transport of ion and fluid in aqueous solutions.[16, 17, 39] Table S2 lists all the boundary conditions used in the simulations.

$$\varepsilon \nabla^2 \varphi = -\sum_{i=1}^{N} z_i F C_i \tag{1}$$

$$\nabla \cdot \mathbf{J}_i = \nabla \cdot \left( C_i \mathbf{u} - D_i \nabla C_i - \frac{F z_i C_i D_i}{RT} \nabla \varphi \right) = 0 \tag{2}$$

$$\mu \nabla^2 \mathbf{u} - \nabla p - \sum_{i=1}^{N} (z_i F C_i) \nabla \varphi = 0 \tag{3}$$

$$\nabla \cdot \mathbf{u} = 0 \tag{4}$$

where $z_i$, $C_i$, $J_i$, and $D_i$ are the valence, concentration, ionic flux, and diffusion coefficient of ionic species $i$ (cations or anions). $u$ is the fluid velocity, and $\varepsilon$ is the dielectric constant of solutions. $\varphi$, $N$, $F$, $R$, $T$, $p$, and $\mu$ are the electric potential, number of ion types, Faraday constant, gas constant, temperature, pressure, and liquid viscosity, respectively.

Ion current through nanopores at different voltages was obtained by integrating the flux of cations and anions at the boundary of the reservoir with Equation 5.[16, 17, 39]

$$I = \int_S F \left( \sum_i^2 z_i \mathbf{J}_i \right) \cdot \mathbf{n} \, dS \tag{5}$$

where $S$ represents the boundary of the reservoir, and $\mathbf{n}$ is the unit normal vector.

To consider the influence of EDLs on the ion transport through nanopores, the mesh size on the inner and exterior (within 3 μm from the pore boundary) pore walls was selected as 0.1 nm. For the other parts of the exterior walls, the mesh size was used as 0.5 nm to lower the calculation cost. The specific mesh construction details are shown in Figure S1.[16, 17]

## Results and Discussion

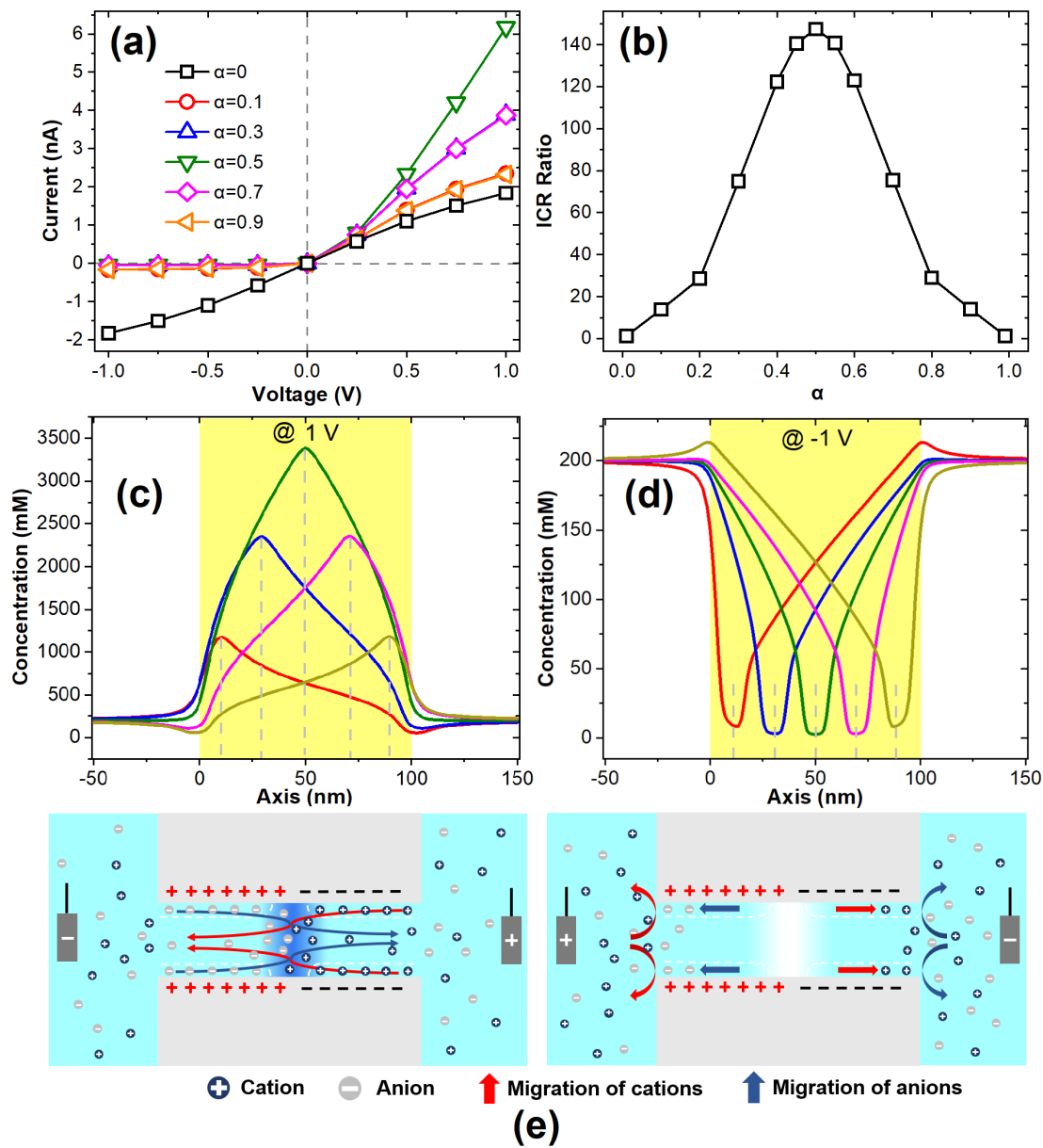

Figure 2 Ionic behaviors in bipolar nanopores with different length ratios of negatively charged surfaces. (a) I-V curves. (b) Ionic current rectification ratios calculating by $|I_{+1V}/I_{-1V}|$. (c-d) Distributions of ion concentration along the pore axis under 1 V (c) and -1 V (d). Vertical grey dashed lines denote the locations of the

junction between differently charged regions. (e) Illustrations of ion migration under ± 1 V.

Nanopores act as an important platform for studying ion transport under confined spaces. When the voltage is applied across the nanopore, free ions in the solution undergo direct migration driven by the electric field, thereby forming an ionic current through the pore. In the ion current circuit, the nanopore works as a resistor, and its resistance value is determined by the solution conductivity, which depends on the salt concentration inside the pore.[47] Ionic current rectification (ICR) has been widely reported with nanopores including conical nanopores, unipolar and bipolar nanopores.[20, 48] The rectification phenomenon under opposite biases is induced from the significantly different statuses of ion accumulation, i.e. ion enrichment for the "on" state and ion depletion for the "off" state.

ICR inside bipolar nanopores originates from the selectivity to counterions of oppositely charged surfaces on both pore ends.[18, 27] For the bipolar nanopore with 10 nm in diameter, the positively and negatively charged ends are selective to the anions and cations, respectively. When cations and anions enter the nanopore selectively, surface charges can provide electrostatic attraction to counterions which can transport rapidly along the EDLs near the charged walls. As counterions approach the junction of positively and negatively charged surfaces, the appearance of oppositely charged surfaces cuts off the passway for ion transport in EDLs. Then, these counterions have to slow down and transport through the samely charged portion in the center region as co-ions. This results in ion enrichment inside the nanopore corresponding to the "on" state in I-V curves. Under the reverse electric field, both cations and anions are repelled from the oppositely charged pore ends which cannot enter the bipolar nanopore. Also, counterions inside the bipolar nanopore flow out of the pore, causing ion depletion due to insufficient ion supplementation.

Since the selectivity of nanopores to counterions is related to the length of the charged region,[13, 16, 17] we investigate the modulation of different charged length ratios of bipolar nanopores on ion transport. In our simulation models, the length ratio of the negatively charged part to the pore length is denoted as α. As the voltage varies from −1 to 1 V, I-V curves were obtained with bipolar nanopores, as shown in Figure 2a. Note that when α equal to 0 or 1, the inner pore walls are completely positively or negatively charged, respectively. For both uniformly charged cases, linear I-V curves are obtained at low voltages. When the voltage exceeds 0.5 V, due to the concentration polarization at both pore ends under strong electric fields,[49] the ion current inside nanopores exhibits the limiting current.[16]

For cases with a nonzero α, positive and negative surface charges carried on inner pore walls form the bipolar charge distribution. As shown in Figure 2a, significant ICR appears in bipolar nanopores, low currents appearing at negative voltages and high currents existing at positive voltages. From the obtained I-V curves, as α gradually increases from 0 to 1, the current value at positive voltages has an increase-decrease profile (Figure S2) which reaches its maximum at α=0.5. At negative voltages, the current exhibits a decrease-increase trend with α, reaching its minimum at α=0.5 (Figure S2).

The ICR ratio is usually considered to characterize the degree of current rectification through nanopores, which is calculated here with the current obtained at ±1 V, i.e. ICR Ratio = $|I_{+1V}/I_{-1V}|$.[13, 48] Figure 2b shows ICR ratios of nanopores with 100 nm in length and 10 nm in diameter under different values of α. When the nanopore has a bipolar charge distribution, it exhibits current rectification with various degrees from ~1 to ~150. The maximum ICR ratio ~150 appears at α =0.5, i.e. the positively charged portion and negatively charged portion have equal length on inner pore walls.

Taking advantage of simulations, detailed ionic behaviors can be obtained inside nanopores. Figures 2c and 2d show the distribution of ion concentration along the pore axis in bipolar nanopores with different length ratios of negatively charged surfaces under ± 1 V, respectively. Under 1 V, nanopores are at the "on" state, corresponding to the ion enrichment inside nanopores. Figure 2e shows the ionic migration through nanopores under opposite voltages. At positive voltages, cations enter the negatively charged pore end near the positive electrode and move toward the positively charged pore end near the negative electrode. Due to the ionic selectivity of charged pore walls, counterions are attracted into the nanopore and transported mainly along the EDLs. When cations reach the positively charged surface, due to the same charge sign, cations are repelled by surface charges and enter the center region of the nanopore. This slows down the transport of cations and results in the enrichment of cations. Similarly, anions entering the pore from the other end also accumulate inside the bipolar nanopore. Note that the most significant ion enrichment happens at the junction between positively and negatively charged surfaces of nanopores. Due to the relatively lower concentration, the higher electric field strength appears at both pore ends which can accelerate the entry process of counterions. Among all cases, bipolar nanopores with half-positive and half-negative surfaces can accommodate the most ions inside.

At negative voltages, both anions and cations are difficult to enter the nanopore due to the electrostatic repulsion from surface charges during the entry process of ions. Also, ions inside the nanopore are quickly discharged under the electric field, leading to the formation of ion depletion. From Figure 2d, the lowest ion concentration appears at the junction of bipolar charge distribution, which corresponds to the highest electric field strength. In this case, the positive feedback can also accelerate the formation of the depletion zone inside bipolar nanopores. Due to the similar depleted situation under various cases with different α, the obtained current shares similar values.

Under the combined influences from ion selectivity and electric field, the ion enrichment and depletion formed inside bipolar nanopores under opposite biases lead to ionic current rectification. At α =0.5, i.e. both positive and negative charges cover half of the pore wall, the most significant ion enrichment and depletion appear inside bipolar nanopores which induce the highest ICR ratios. In Figure 2, due to the similar ion mobility of $K^+$ and $Cl^-$ ions in KCl solutions, in two cases where the two α add up to 1, the ICR ratios share approximately the same values, and the ion concentration in each case shows the mirror-symmetric distribution.

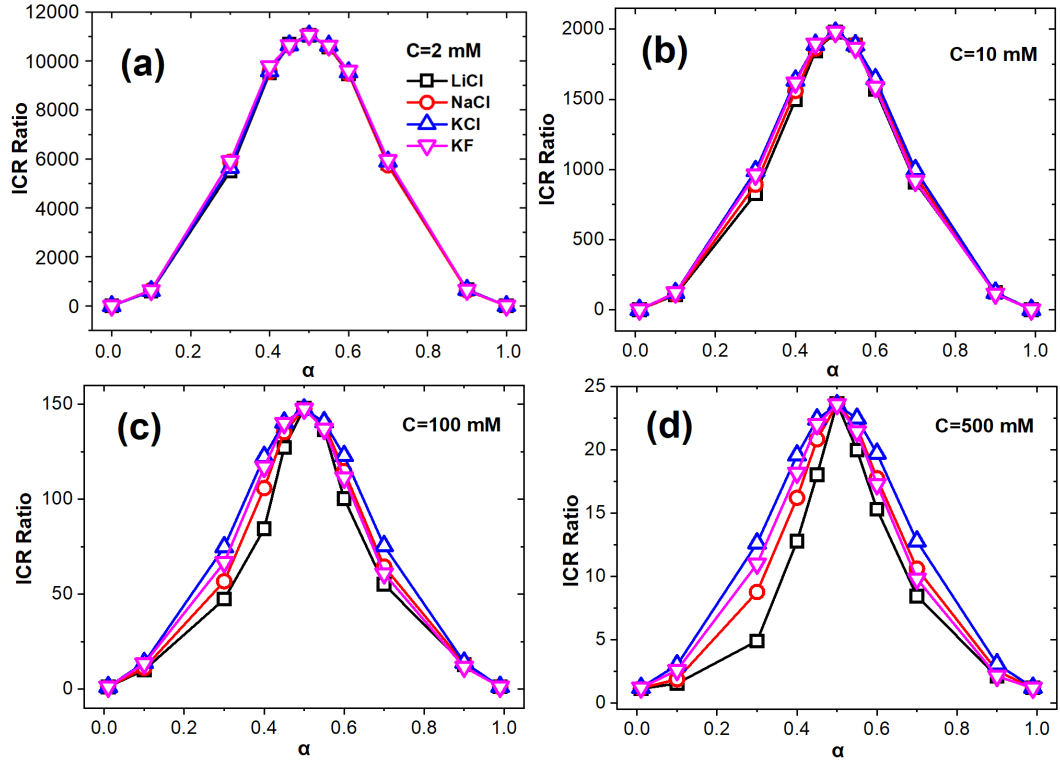

Figure 3 Ion current rectification (ICR) ratios in bipolar nanopores under various concentrations and salt types. (a) C=2 mM. (b) C=10 mM. (c) C=100 mM. (d) C=500 mM. The diameter and length of nanopores were 10 nm and 100 nm, respectively. The surface charge density was −0.08 C/m$^2$.

Due to the dependence of ionic transport on the properties of aqueous solutions,[2] we investigated the modulation of ICR in bipolar nanopores from the salt concentration and type with different charge length ratios. As shown in Figure 3, different salt types, such as LiCl, NaCl, KCl, and KF, as well as bulk concentrations from 2 to 500 mM are considered to explore the influences of diffusion coefficients of cations and anions and the Debye length on ICR ratios. In 2 mM solutions, a Debye length of ~6.8 nm is yielded which induces overlapped EDLs inside nanopores.[2, 44] As α changes from 0 to 1, ICR ratios in the four kinds of solutions exhibit almost the same increase-decrease profiles, which share an equal maximum value at α=0.5. It indicates that the ICR ratio through bipolar nanopores with half-positive and half-negative surfaces is independent of the ionic mobility of cations or anions. In the cases with a higher concentration of 10, 100, or 500 mM, a similar trend of the ICR ratio with decreased values varying with α is obtained. The maximum ICR ratio persists at α=0.5.

In a pair of bipolar nanopores with two α values with addition equals 1, due to the difference between diffusion coefficients of cations and anions, ICR ratios may have large deviations. In KCl solutions, ICR ratios exhibit a good symmetry with α at various concentrations, which is due to the similar diffusion coefficients of $K^+$ and $Cl^-$ ions. In LiCl solutions, due to the much smaller diffusion coefficient of $Li^+$ ions than $Cl^-$ ions, the symmetry of its ICR ratio profile is weak. With the concentration increasing, the asymmetry becomes more significant. For bipolar nanopores with unequal lengths of positive and negative surfaces, the ICR ratios show dependence on the relative ion mobilities. In such aqueous solutions, with a larger difference in the diffusion coefficient of cations and anions, smaller ICR ratios are induced. As the salt concentration increases, the deviation among ICR ratios in considered salt types becomes more obvious.

Considering that bipolar nanopores with half-positive and half-negative surfaces have the most significant ICR ratio, which is independent of the ion type, α is selected to be 0.5 in the subsequent study.

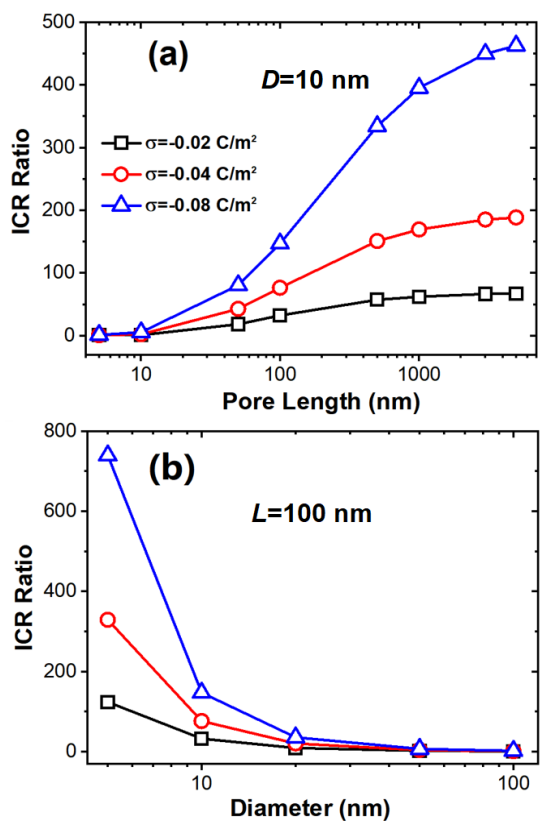

Figure 4 ICR ratios through bipolar nanopores under various nanopore parameters, such as pore lengths (a), and pore diameters (b). In simulations, 100 mM KCl was used. For cases in (b) the surface charge density was −0.08 C/m$^2$.

Properties of the nanopore include its geometric dimensions and surface charge density. Here, the modulation of ICR ratios in bipolar nanopores with α=0.5 is investigated by adjusting the pore length and diameter, as well as the surface charge densities. From Figure 4a, for 10-nm-diameter nanopores with a surface charge density of −0.02 C/m$^2$, as the pore length increases from 10 to 1000 nm, the ICR ratio enhances from ~1 to ~54. In long nanopores, due to the stronger ionic selectivity, more significant ion enrichment and depletion are induced inside nanopores. With the pore length further increasing to 5 μm, the ICR ratio gradually approaches a saturation of ~60. The strength of surface charges plays a decisive role in the ionic selectivity of nanopores,[12] which also has a significant influence on ICR ratios.[18] With the surface charge densities increasing from −0.02 to −0.04 or −0.08 C/m$^2$, ICR ratios under the same conditions can be enhanced by 3 or 8 times, respectively. This is attributed to the higher degree of ion enrichment and depletion under opposite voltage biases due to the modulation of ion transport by stronger surface charges.

The pore diameter controls the degree of confinement in nanopores which has significant influences on the ion transport in the presence of surface charges. Figure 4b shows the dependence of ICR ratios on the pore diameter at the pore length of 100 nm. As the pore diameter expands, the modulation of surface charges on ion transport gets weakened, leading to a lower degree of ion enrichment and depletion and decreased ICR ratios. When the pore diameter increases to 100 nm, the ICR ratio decreases to ~1 which indicates that the bipolar nanopore loses the current rectification.

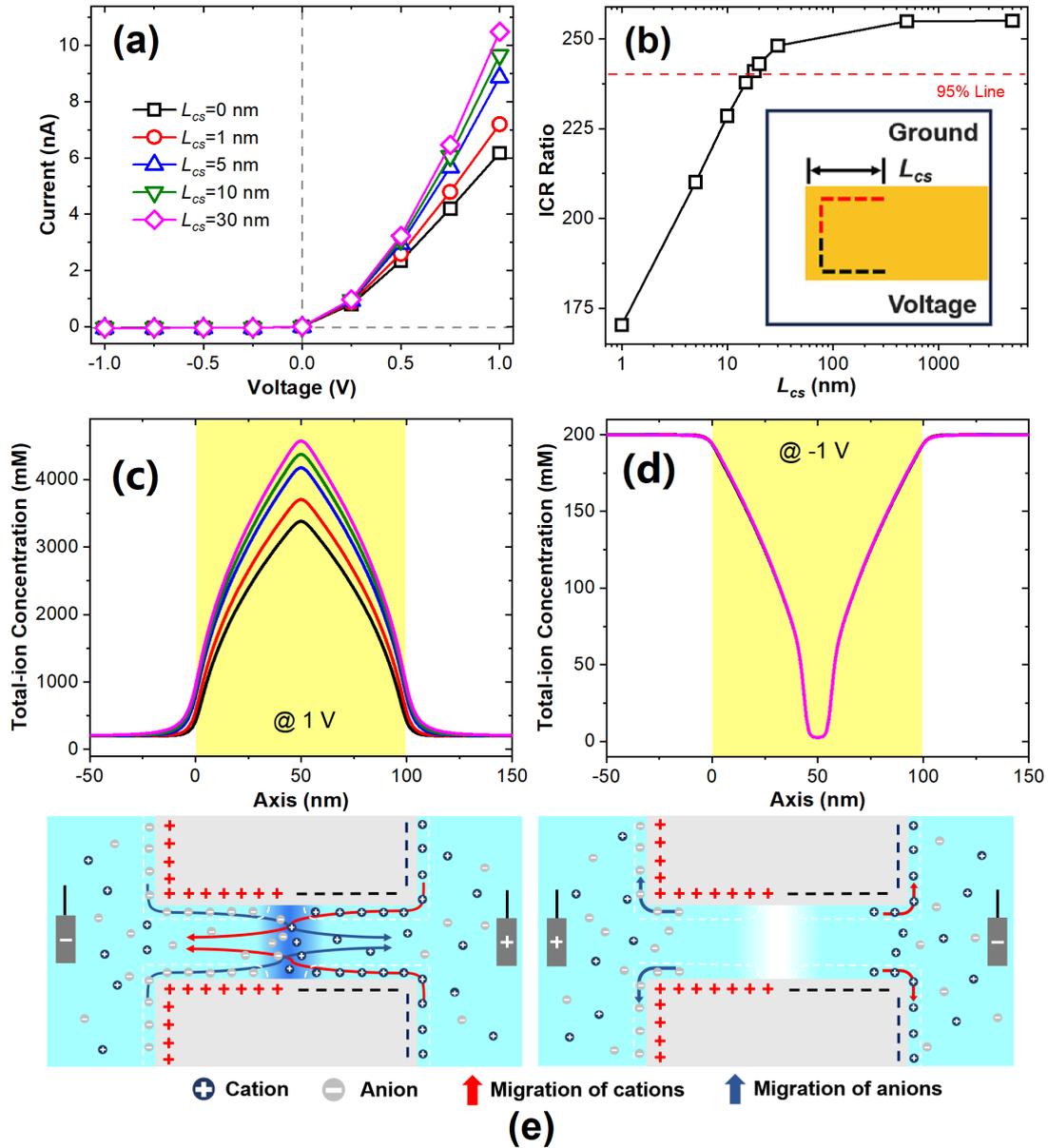

Figure 5 Ionic behaviors in bipolar nanopores affected by the charged exterior surfaces. Charged exterior surfaces are considered as a ring region with a width of $L_{cs}$. (a) I-V curves. (b) Ionic current rectification ratios. (c-d) Distributions of ion concentration along the pore axis under 1 V (c) and −1 V (d). (e) Illustrations of ion migration under ± 1 V.

On membranes, as a 3D structure, the nanopore has both inner and external surfaces. From our previous work,[13, 16, 17] for nanopores with sub-200 nm lengths, external surface charges on the membrane can provide significant influences on the transport of ions and fluid through nanopores under electric fields or salt gradients. From the inset of Figure 5b, the width of the charged area beyond the pore boundary is denoted as $L_{cs}$. Note that at $L_{cs}$=0, the bipolar nanopore has no external surface charges.

Simulations containing a 100-nm-length and 10-nm-diameter bipolar nanopore were conducted with the $L_{cs}$ increasing from 0 to 5 μm to explore the influence of the charged external area on the ICR ratio. Under different $L_{cs}$, I-V curves through bipolar nanopores are collected in Figure 5a. Without external surface charges, i.e. $L_{cs}$=0, the nanopore stays at the "on" and "off" states under positive and negative voltages, respectively. After the external surfaces get charged, the ionic current at the "on" state enhances, while that at the "off" state remains almost constant. With $L_{cs}$ increasing from 0 to 30 nm, the current value at 1 V is promoted from ~6.2 to ~10.5 nA, increasing by 69%. As $L_{cs}$ expands further, the current gradually reaches its saturation.

ICR ratios under various $L_{cs}$ are shown in Figure 5b. With $L_{cs}$ increasing from 0 to ~20 nm, the ICR ratio gets enhanced from 147 to 243, increasing by 65%. This is mainly attributed to that the appearance of external surface charges can enhance the ion enrichment inside nanopores (Figures 5c and 5d), corresponding to the increased maximum value in the distribution of ion concentration along the pore axis (Figure S3). As illustrated in Figure 5e, EDLs formed near charged exterior surfaces can form a high-concentration pool of counterions and provide a fast passageway for the migration of counterions.[16] The larger supplementation of counterions induces a higher degree of ion enrichment. However, at negative voltages, the degree of ion depletion is almost unaffected by the $L_{cs}$ (Figure 5d).

As $L_{cs}$ further increases, the ICR ratio increases slowly and gradually approaches its saturation.[33] The effective width of the charged exterior surface ($L_{cs\_eff}$) is defined as the value of $L_{cs}$ at which the ICR ratio reaches 95% of that at $L_{cs}$=5000 nm.[13, 16, 17] For the nanopore with a length of 100 nm, a diameter of 10 nm, and a surface charge density of −0.08 C/m$^2$, the $L_{cs\_eff}$ is ~20 nm.

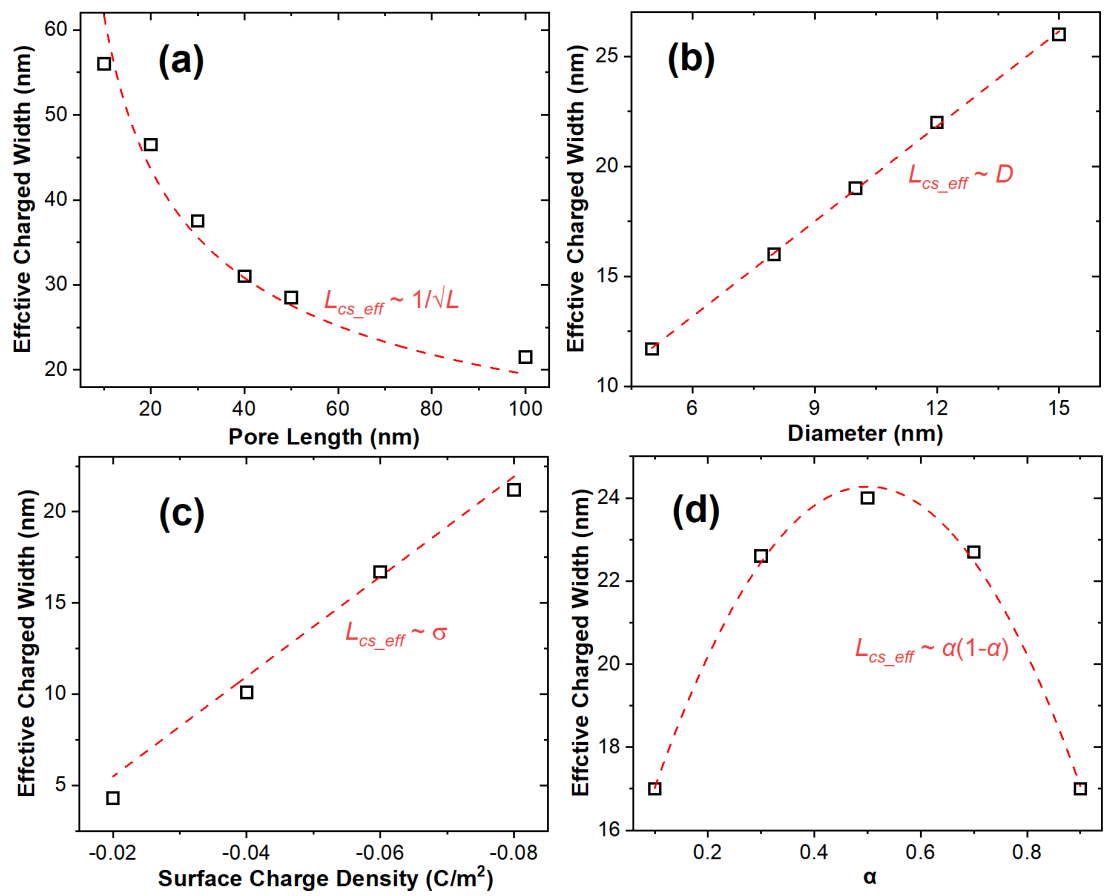

Figure 6 Effective charged width beyond the pore boundary ($L_{cs\_eff}$) under various nanopore parameters, such as pore length (a), diameter (b), surface charge density (c), and α (d).

The effective charged width for individual nanopores is important in propagating the investigation of ion transport in individual nanopores to porous membranes.[16, 31, 34] With a larger separation between individual nanopores than the effective charged width, each nanopore can function independently without any interference among nanopores. The effective charged width can serve as an important parameter in the design of functional porous membranes in various applications, such as osmotic energy conversion,[34, 36] and parallel nanofluidic sensing.[50, 51]

From the above simulation results, charged exterior surfaces have a significant impact on the ion transport inside nanopores. Here, a series of simulations have been conducted with the consideration of various pore properties such as dimensions, surface charge density, and length ratio α, and external conditions, like the applied voltage, salt concentration, and salt type. We systematically collected the effective charged length under different conditions and explored the quantitative relationship between the effective length of charged exterior surfaces and these parameters.

From Figure 6a, in cases under different pore lengths, $L_{cs\_eff}$ decreases inversely proportional to the square root of the pore length. The increase in pore lengths can lead to an increase in the resistance inside nanopores. Then, voltages at both pore ends are reduced, which weakens the ion transport along charged external pore walls.

Similar to nanopores with pure positive or negative surfaces, the modulation of charged exterior surfaces on ion transport can be neglected when the pore length is long enough.

In Figure 6b, $L_{cs\_eff}$ has a linear dependence on the pore diameter. With the diameter expanding from 5 to 15 nm, $L_{cs\_eff}$ is enlarged from 11 to 26 nm. Due to the linear dependence of the inner-pore wall area on the pore diameter,[17] more counterions can be transported along charged pore walls. Then, larger charged exterior surfaces are required accordingly to supply more ions towards pore entrances along charged external surfaces from the bulk far away from the orifice.

The enhancement of the surface charge density induces a larger aggregation of counterions inside EDLs, which thereby promotes ionic selectivity as well as ion enhancement under positive voltages. The increased ion flux through the nanopore requires a larger supplementation of counterions from farther locations away from the orifices, which can be transported to both pore entrances by the external surface conductance.[16] Therefore, $L_{cs\_eff}$ increases linearly with the strength of surface charge density.

The effect of the surface charge part on $L_{cs\_eff}$ was also considered as shown in Figure 6d. A symmetrical relationship is found between α and the effective length of charged exterior surfaces, whose axis is located at α=0.5. In two simulation cases with both individual α values adding to 1, the obtained $L_{cs\_eff}$ share almost the same values. This is due to the similar diffusion coefficients of anions and cations in KCl solutions. As α approaches 0.5, the greater degree of ion enrichment and depletion inside nanopores leads to a larger influence of exterior surface charges on ion transport, which results in the increased effective length of charged exterior surfaces.

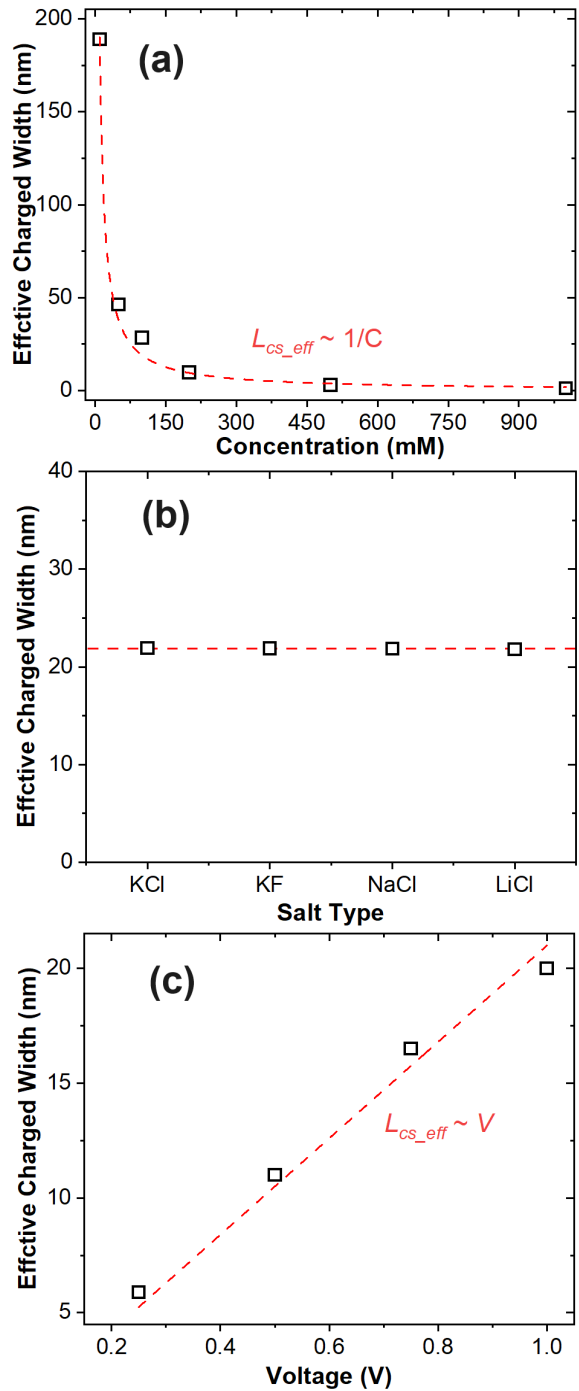

Figure 7 Effective charged width beyond the pore boundary ($L_{cs\_eff}$) under various simulation conditions, such as the salt concentration (a), salt type (b), and applied voltage (c).

Figure 7 shows the influences of simulation conditions on $L_{cs\_eff}$ including the solution parameters and applied voltages. In nanofluidic systems, the salt concentration determines the Debye length, which controls the degree of surface charge screening by counterions.[44] In a higher-concentration solution, counterions can shield surface charges better, reducing the modulation of surface charges on ion transport through nanopores. Similar to the inverse relationship between $L_{cs\_eff}$ and the salt concentration found in our previous investigations with uniformly charged nanopores,[16] as shown in Figure 7a, $L_{cs\_eff}$ inversely correlates to the salt concentration with bipolar nanopores. With the salt concentration increasing from 10 to 200 mM, $L_{cs\_eff}$ decreased significantly from 189 to 9.8 nm. As the concentration further increases from 200 to 1000 mM, $L_{cs\_eff}$ changes from 9.8 to 1.2 nm exhibiting a weak dependence on the salt concentration, due to the weak effect of surface charges on ion transport through nanopores.[2]

Figure 7b presents the dependence of the effective length of charged exterior surfaces on salt types that possess different ion diffusion coefficients. For the four types of electrolytes considered in this work, the $L_{cs\_eff}$ is independent of the ion diffusion coefficient. This may be due to the almost same degrees of ion enrichment and depletion inside bipolar nanopores with half-positive and half-negative surfaces.

The relationship between the applied voltage and $L_{cs\_eff}$ is shown in Figure 7c. Under a stronger electric field, the transport of a larger number of ions is required to form the more serious ion enrichment and depletion inside nanopores. Due to the dependence of ion enrichment on $L_{cs\_eff}$, the large number of counterions in the bulk needed to supplement by external surface conductance induces a larger $L_{cs\_eff}$ at a higher voltage.

From our data shown in Figures 6 and 7, $L_{cs\_eff}$ depends closely on various pore properties, as well as solution concentration and applied voltages, which may induce a complex analytical description for $L_{cs\_eff}$. Considering the convenient usage in practical applications, the analytical description of $L_{cs\_eff}$ will be focused on in the short future after the detailed theoretical prediction of ion transport is obtained as shown in previous works.[52-55]

## Conclusions

Due to the asymmetric surface charge distribution, bipolar nanopores exhibit significant ionic current rectification under equal voltages but opposite polarities. In this work, the modulation mechanism of ICR in bipolar nanopores was explored with simulations via the adjustment of the length ratio (α) of negative surfaces. At the "on" state, as α gradually increases from 0 to 1, the current value has an increase-decrease profile. At α =0.5, i.e. bipolar nanopores with half-positive and half-negative surfaces, the ICR ratio can reach its highest value of ~150 due to the most significant ion enrichment and depletion inside the nanopore. In simulations with bipolar nanopores of different dimensions and surface charge properties, the ICR becomes more obvious with the increase of pore length which then reaches the saturation. The saturated ICR ratio is proportional positively to the surface charge density. Charged exterior surfaces of nanopores can induce enhanced ion enrichment, which increases the ICR ratio from 147 to 243 with the width of charged exterior surface $L_{cs}$ increasing from 0 to ~20 nm. As $L_{cs}$ increases further, the ICR ratio gradually approaches its saturation. Then the effective area of the charged exterior surface near nanopores was investigated under various simulation conditions including the pore dimension, surface charge properties, applied voltages, and solution parameters. For a 100-nm-long and 10-nm-diameter bipolar nanopore with half-positive and half-negative surfaces of 0.08 C/m$^2$, the effective $L_{cs}$ is ~20 nm. The effective $L_{cs}$ has a linear relationship with the pore diameter, surface charge density, and applied voltage, a reciprocal relationship with the pore length and salt concentration, as well as a symmetrical relationship with the charged length ratio α. Also, the effective $L_{cs}$ is independent of salt type. Our simulations in this work provide the microscopic view of ion transport through bipolar nanopores. The results can guide the design of bipolar-charged porous films for nanofluidic devices, energy conversion, and related applications.

## Conflicts of interest

The authors declare no competing financial interest.

## Acknowledgment

This research was supported by the National Natural Science Foundation of China (52105579), the Guangdong Basic and Applied Basic Research Foundation (2023A1515012931), the Instrument Improvement Funds of Shandong University Public Technology Platform (ts20230107), Key Research and Development Program of Yancheng (BE2023010), and the Qilu Talented Young Scholar Program of Shandong University.

## Supporting Information

See the Supporting Information for surface potentials, detailed boundary conditions and mesh strategy used in our simulations, variation of ionic current with α at ±1 V, and the maximum value of concentration profiles along the nanopore axis.

# Modulation of ionic current rectification in short bipolar nanopores

Hongwen Zhang,[1,2] Long Ma,[1] Chao Zhang,[3] and Yinghua Qiu[1,2,4*]

1. Shenzhen Research Institute of Shandong University, Shenzhen, 518000, China

2. Key Laboratory of High Efficiency and Clean Mechanical Manufacture of Ministry of Education, National Demonstration Center for Experimental Mechanical Engineering Education, School of Mechanical Engineering, Shandong University, Jinan, 250061, China

3. School of Mechanical and Electronic Engineering, Shandong Jianzhu University, Jinan, 250101, China

4. Suzhou Research Institute of Shandong University, Suzhou, 215123, China

*Corresponding author: yinghua.qiu@sdu.edu.cn

1. Surface potential on charged surfaces

Table S1 Corresponding surface potential in considered cases with different surface charge densities at various concentrations. All values were obtained in 1:1 electrolyte solutions.

| Surface charge density (C/m$^2$) | Concentration (mM) | Surface potential obtained by Grahame equation (mV) | Surface potential obtained from simulations (mV) |
|---|---|---|---|
| −0.02 | 100 | -26.58 | -28.90 |
| −0.04 | 100 | -48.19 | -51.34 |
| −0.06 | 100 | -64.79 | -68.17 |
| −0.08 | 100 | -77.82 | -81.42 |
| −0.08 | 2 | -175.86 | -182.89 |
| −0.08 | 10 | -134.71 | -140.84 |
| −0.08 | 500 | -44.09 | -45.38 |
| −0.08 | 800 | -36.22 | -37.03 |
| −0.08 | 1000 | -32.86 | -33.49 |

The Grahame equation is shown below(Eq. S1).

$$\sigma = \sqrt{8k_b T C \varepsilon \varepsilon_0} \sinh\left(\frac{e\varphi_0}{2k_b T}\right) \quad (S1)$$

where $\sigma$ and $\varphi_0$ are the surface charge density and the surface potential. $K_b$, and $T$ are the Boltzmann constant and temperature. $C$ is the salt concentration. $\varepsilon$ and $\varepsilon_0$ are the dielectric constants of water and vacuum.

2. Simulations details.

Table S2 Boundary conditions used in simulations. Coupled Poisson–Nernst–Planck and Navier–Stokes equations were solved with COMSOL Multiphysics.

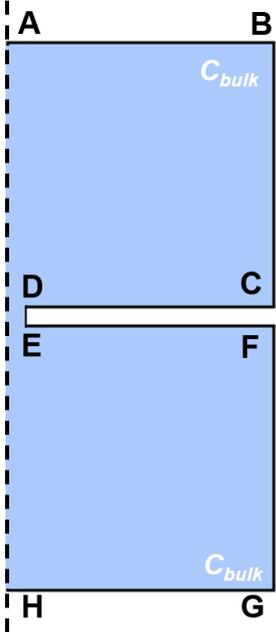

| Scheme | Surface | Poisson | Nernst-Planck | Navier-Stokes |
|---|---|---|---|---|
| | AB | Constant potential (Ground) $\phi = 0$ | Constant concentration $c_i = C_{bulk}$ | Constant pressure $p = 0$ No viscous stress $\mathbf{n}\cdot[\mu(\nabla\mathbf{u}+(\nabla\mathbf{u})^T)]=0$ |
| | BC, FG | Neutral surface $-\mathbf{n}\cdot(\varepsilon\nabla\phi)=0$ | No flux $\mathbf{n}\cdot\mathbf{N}_i = 0$ | No slip $\mathbf{u} = 0$ |
| | CD, DE, EF | $-\mathbf{n}\cdot(\varepsilon\nabla\phi)=\sigma_w$ | No flux $\mathbf{n}\cdot\mathbf{N}_i = 0$ | No slip $\mathbf{u} = 0$ |
| | HG | Constant potential $\phi = V$ | Constant concentration $c_i = C_{bulk}$ | Constant pressure $p = 0$ No viscous stress $\mathbf{n}\cdot[\mu(\nabla\mathbf{u}+(\nabla\mathbf{u})^T)]=0$ |
| | AH | Axial symmetry | Axial symmetry | Axial symmetry |

$\phi$, $V$, $\varepsilon$, $\sigma_w$, $C_{bulk}$, $p$, $\mathbf{n}$, $\mathbf{N}_i$, $\mu$, $\mathbf{u}$ are the surface potential, applied voltage, dielectric constant, surface charge density of the pore wall, bulk concentration, pressure, normal vector, flux of ions, solution viscosity and fluid velocity, respectively.

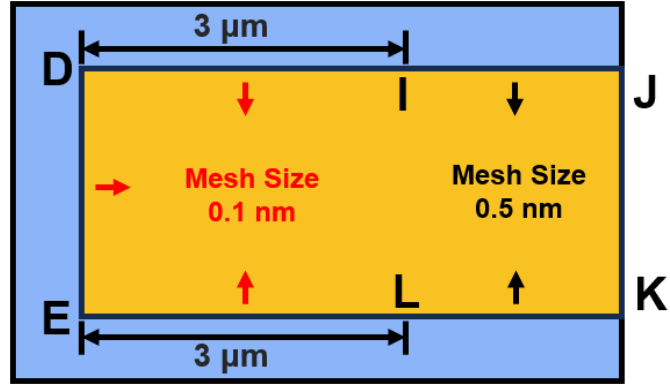

Figure S1 Mesh strategy used in the simulations. The mesh size on the inner-pore surface (DE) and 3 μm-wide regions of exterior surfaces beyond the pore boundary (DI and EL) was selected as 0.1 nm. For the other parts of the exterior walls (IJ and LK), the mesh size was used as 0.5 nm.

3. Additional simulation data.

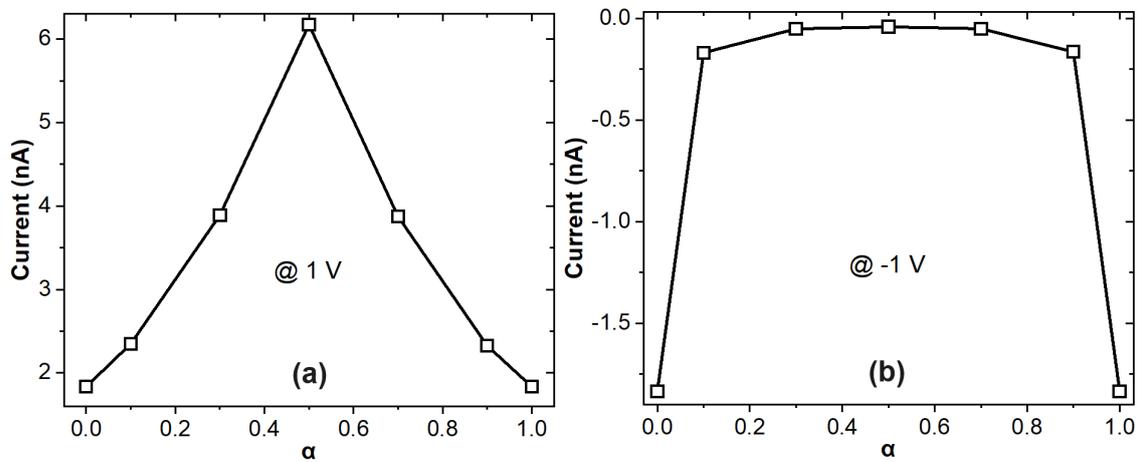

Figure S2 Variation of ionic current with α at ±1 V. The pore length and diameter were 100 and 10 nm. The length ratio of the negatively charged part to the pore length α was varying from 0 to 1. The value of the surface charge density was −0.08 C/m². Solutions were 0.1 M KCl. The voltage of 1 V (a) and −1 V (b) were applied across the nanopore respectively.

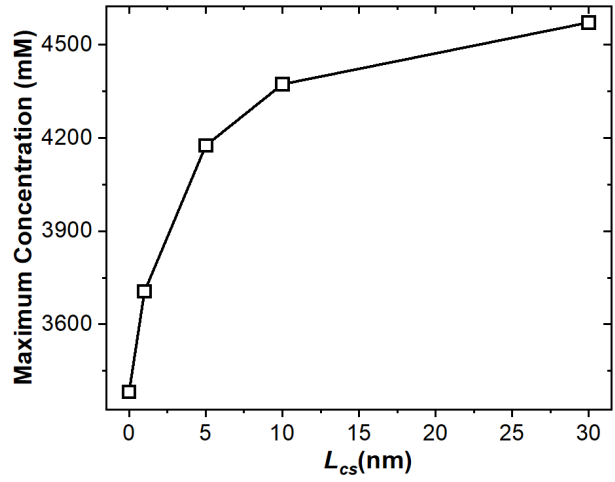

Figure S3 The maximum value of concentration profiles along the nanopore axis under different $L_{cs}$ varying from 0 to 30 nm. The pore length and diameter were 100 and 10 nm. The length ratio of the negatively charged part to the pore length α was 0.5. The value of the surface charge density was −0.08 C/m². Solutions were 0.1 M KCl. The voltage of 1 V was applied across the nanopore.